\documentclass{aastex631}
\usepackage{tikz}
\usetikzlibrary{arrows.meta,bending}

\usepackage{latexsym}
\usepackage{lmodern}

\usepackage{amsmath,amsfonts,amssymb,amsthm,amstext,amscd,mathtools,eucal,esint,color}

\usepackage[all]{xy}
\usepackage{dsfont,epiolmec,comment}
\usepackage{physics}
\usepackage{mathrsfs,calligra}
\usepackage{yfonts}
\usepackage{subfigure}
\usepackage{hyperref,comment}

\def\L{{\mathcal L}}

\newcommand{\be}{\begin{equation}}
\newcommand{\ee}{\end{equation}}
\newcommand{\bea}{\begin{eqnarray}}
\newcommand{\eea}{\end{eqnarray}}

\newcommand{\bsb}{\boldsymbol}

\def\e{\mathrm{e}}

\def\T{\mathcal{T}}
\def\d{\partial}
\def\q{{\bsb q}}
\def\v{{\bsb v}}
\def\p{{\bsb p}}
\def\k{{\bsb k}}
\def\Q{{\bsb q}}
\def\K{{\bsb k}}
\def\x{{\bsb x}}
\def\O{\mathcal{O}}
\def\cH{\mathcal{H}}
\def\y{{\bsb y}}
\def\s{{\bsb s}}

\newcommand{\gsim}{\lower.7ex\hbox{$\;\stackrel{\textstyle>}{\sim}\;$}}
\newcommand{\lsim}{\lower.7ex\hbox{$\;\stackrel{\textstyle<}{\sim}\;$}}

\usepackage{colortbl}
\definecolor{summersky}{cmyk}{0.71,0.33,0,0.14}
\definecolor{flamingo}{cmyk}{0,0.51,0.71,0.14}
\definecolor{rp}{cmyk}{0.2, 1, 0.6, 0}
\definecolor{pacificblue}{cmyk}{0.95,0.3,0, 0.19}
\definecolor{gray60}{cmyk}{0.4,0.4,0,0.8}
\definecolor{green94}{cmyk}{94,0,100,0}
\definecolor{green80}{cmyk}{80,0,90,0}

\definecolor{darkgreen}{rgb}{0,0.3,0}
\definecolor{darkblue}{rgb}{0,0,0.3}
\definecolor{darkred}{rgb}{0.7,0,0}

\definecolor{summersky}{cmyk}{0.71,0.33,0,0.14}
\definecolor{flamingo}{cmyk}{0,0.51,0.71,0.14}

\graphicspath{{./}{figures/}}
\usepackage[section]{placeins}
\usepackage{float}

\begin{document}

\title{The Effective Field Theory of Large Scale Structures of a Fuzzy Dark Matter Universe}

\author[0000-0001-8233-9590]{Hamed Manouchehri Kousha}\altaffiliation{	hamed.manouchehri@physics.sharif.edu}
\affiliation{Department of Physics, Sharif University of Technology, Tehran, Iran}

\author[0000-0002-3142-7168]{Sina Hooshangi}\altaffiliation{sina.hooshangi@ipm.ir}
\affiliation{School of Astronomy, Institute for Research in Fundamental Sciences (IPM), Tehran, Iran}

\author[0000-0002-4442-1523]{Aliakbar Abolhasani}\altaffiliation{ali.abolhasani@sharif.edu}
\affiliation{Department of Physics, Sharif University of Technology, Tehran, Iran}

%\linenumbers

\begin{abstract}
    Ultra-light scalar fields and their non-interacting class, i.e., the so-called fuzzy dark matter (FDM), are dark matter candidates introduced to solve the small-scale problems of the standard cold dark matter. In this paper, we investigate whether the physics of FDM, particularly the quantum pressure that leads to the suppression of structure formation on small scales, could leave significant imprints on the large-scale statistics of matter fluctuations. For this purpose, we utilize the Effective Field Theory of Large Scale Structures (EFT of LSS), wherein small-scale physics is integrated and represented on large scales by only a set of free parameters. These parameters can be determined by fitting them into the cosmological simulations. By fitting the EFT predictions to the simulation data, we determine the value of the speed of sound as a quantitative measure of how UV physics affects large-scale perturbation. We use the \textit{Gadget-2} code to study the evolution of $512^3$ particles in a box of side length $250\,h^{-1}\,\mathrm{Mpc}$. We exploit the suppressed FDM initial power for the FDM universe and perform N-body simulation sufficient to produce accurate - enough for our purpose - results on large scales. Particularly, we perform three FDM simulations with different masses and compare their sound speed with the standard cold dark matter (CDM) simulation. We found no difference between the FDM and CDM sound speeds beyond the confidence intervals. However, a consistently increasing trend can be seen in the sound speed for lower masses. This result suggests further investigations using higher-resolution simulations.

\end{abstract}

\section{Introduction}
According to the standard model of cosmology, nearly $26$ percent of the universe's energy content consists of some cold matter with negligible non-gravitational interaction, called \textit{dark matter} \citep{Planck}. The standard candidate for dark matter is weakly interacting massive particles (WIMPs). The theoretical predictions based on WIMPs are consistent with large-scale observational data. However, in small scales ($\sim 10\, \mathrm{kpc}$), some discrepancies emerge, e.g., the core-cusp problem \citep{core-cusp}, the missing satellite problem \citep{missing}, and the too-big-to-fail problem \citep{toobig}. There are two main ways in which people try to resolve these problems: either by exploring baryonic feedbacks such as supernova explosions that might be responsible for the disruption of small-scale structures or by proposing other dark matter candidates with new physics on small scales.

One of the alternative candidates for dark matter is ultra-light scalar fields \citep[ULSFs, ][]{Hu_2000}. With a mass of about $10^{-22}\, \mathrm{eV}$, they have a long de-Broglie wavelength showing wave effects\footnote{Usually, these effects are referred to as "quantum effects," while due to the very high occupancy number, the quantum fluctuations are negligible, and the system is classical. Therefore, we prefer the term "wave effects." \citep[see ][for a discussion]{HuiRev}} on galactic scales, which could resolve the small-scale problems \citep[see e.g.][for review]{ferreira2021ultralight,HuiRev,Urena-Lopez:2019kud}. Strictly speaking, the uncertainty principle appears as an additional pressure term in Euler's equation, the so-called \textit{quantum pressure} (QP). It results in smooth cores in the center of halos rather than sharp cusps, which is the prediction of the standard CDM. The QP also tends to smear small-scale non-linearities. The suppression of the amplitude of the perturbations, in turn, leads to a fall-off of the matter power spectrum on small scales, which means that fewer low-mass halos and sub-halos will form. This could solve the missing satellite problem. This suppression, along with the lower maximum circular speed of the baryons in FDM halos, could also relieve the "too big to fail" problem. However, these are still controversial \citep[see e.g.][]{namjoo,helpingor}, and there is no consensus on whether baryonic feedback or alternative candidates like FDM could resolve the small-scale problems completely in a coherent manner \citep[see e.g.][for review]{witten,smallscale1,smallscale2}.

In addition to the suppression of small-scale structure formation, ULSFs have some other fingerprints, including the formation of some bound objects at the center of halos, due to the balance of gravity and QP, i.e., the so-called \textit{solitonic cores}, or the formation of \textit{wave interference patterns} of the size of the de-Broglie wavelength. In general, ULSFs could also have self-interactions. The simplest class of ULSF dark matter, which has no self-interactions, is often called "Fuzzy Dark Matter" \citep[FDM, see, e.g.][for a recent review]{Hui}.

In recent years, cosmological simulations based on FDM dynamics have been widely used to study interference patterns \citep[e.g.][]{interference, Hui}, the merger of solitonic cores \citep[e.g.][]{merger, pyultraligh}, suppression of the mass power spectrum \citep[e.g.][]{Hui,ax-gadget,Simon}, suppression of the halo mass function \citep[e.g.][]{schive2016, Simon, zhanglym}, mixed fuzzy and cold dark matter \citep[e.g.][]{axionyx}, oscillations and random walks of the solitonic cores \citep[e.g.][]{Huirandom, schiverandom}, etc. Resolving the new physics of FDM at the de-Broglie wavelength scale in a large-box simulation requires substantial computational resources. Hence, most previous studies have used relatively small-box simulations to study the small-scale new physics of FDM. As a result, the possible effects of new UV physics on the large-scale dynamics of the universe have not been addressed yet; that needs FDM simulations with a box size of at least $\sim 200\,h^{-1}\,\mathrm{Mpc}$. The study of FDM structure formation at cosmological scales is the subject of this work. Specifically, to circumvent the challenge mentioned above, we use the modified initial power spectrum in an ordinary N-body simulation to achieve this goal.

The universe, at its largest scales, is almost homogeneous with tiny fluctuations. Hence, its dynamics are amenable to the perturbation theory. However, as we approach the scale of clusters and galaxies, the Universe becomes clumpy and non-linear. The Effective Field Theory of large-scale structures (EFT of LSS) is a framework for the study of matter perturbations in linear and quasi-linear regimes \citep{Baumann:2010tm,Carrasco:2012cv,Hertzberg_2014,twoloops,irsafe,irresummed,bispectrum,allredshifts,precision,Abolhasani_2016}. The primary mission of the EFT of LSS is to push the range of validity of the perturbation theory toward the non-linear regime. As long as we are interested in the dynamics of the large-scale perturbations, we can integrate short modes at the expense of appearing as effective sources on the right-hand side of the fluid equations. While the EFT of LSS fixes the general form of these source terms, it cannot say anything about the actual values of the parameters of the effective fluid, particularly the speed of sound or the bulk viscosity. To determine these parameters, one has to resort to large-box computer simulations \citep{Baumann:2010tm}. Different cosmological parameters or dynamics on large scales could change the values of these parameters; however, it is not clear whether new physics on small scales could also change them. This work is to answer this question in the case of FDM. Specifically, we study the impact of the special UV physics of FDM, i.e., the QP, on the one-loop speed of sound parameter of EFT of LSS, in comparison to the standard cold dark matter(CDM). This parameter has been determined using large-box CDM simulations in \cite{irresummed},\cite{allredshifts}, and \cite{precision}. We use the same procedure for large-box CDM and FDM simulations, performed using the public \textit{Gadget-2}\footnote{\url{ https://wwwmpa.mpa-garching.mpg.de/gadget/} } \citep{Gadget-2} code with the proper initial conditions for each, and compare the results.

The paper is organized as follows: In Sec. \ref{sec:SPT}, we briefly review the standard perturbation theory (SPT) and highlight some relevant essential points. In Sec. \ref{sec:FDM}, we discuss the main ideas of FDM, including its dynamical equations and the FDM perturbation theory. The subject of Sec. \ref{sec:sim} is to explain the details of our cosmological simulations of CDM and FDM. Finally, in Sec. \ref{sec:EFT}, after explaining our procedure for determining the EFT parameters of CDM and FDM using our simulations, we discuss and compare some of the main results.

\section{Standard Perturbation Theory} \label{sec:SPT}
Let us first briefly review the main lines of the SPT \citep[see][for a comprehensive review]{BernardeauSPT}.
The equations governing the evolution of the matter density contrast, $\delta$, and the matter velocity field, $\v$, within the SPT are\footnote{As usual, all scales are comoving throughout this paper, unless explicitly stated otherwise.}

\begin{equation} \begin{aligned}
\label{sptcoupled}
\delta' +\theta = &-\d_i(\delta~ v^i)\,,
\end{aligned} \end{equation} \begin{equation} \begin{aligned}
\v' + \cH \v +\nabla \phi = & -\v\cdot \nabla \v,
\end{aligned} \end{equation}
where $\phi$ is the gravitational field, $\theta$ is the divergence of velocity field, $\theta \equiv \nabla. \v$, The primes denote the derivative with respect to the conformal time $\tau$, such that $\partial /\partial \tau\equiv a\, \partial/\partial t$, where $a$ represents the scale factor. The conformal expansion rate is represented by $\cH$, which is equal to $aH$. The solution of this coupled system of equations is usually presented in the form of the following perturbation expansion 
\begin{align}
\label{eq:perturb-exp}
\delta(\k,\tau)=\sum_{n=1}^{\infty} a(\tau)^n\,\delta_n(\k),\qquad    \theta(\k,\tau)=-{\cal H}(\tau)\,\sum_{n=1}^{\infty} a(\tau)^n\,\theta_n(\k),
\end{align}
where $\delta_n(\k)$ and $\theta_n(\k)$ are written as a product of initial values of the fields $\delta$ and $\theta$ integrated against the so-called SPT kernels as
\begin{equation} \begin{aligned}
& \delta_{n} (\boldsymbol{k})= \int_{q_1} \dotsi \int_{q_n} (2\pi)^3 \delta^3(\boldsymbol{k} - \boldsymbol{q}_1 -\dotsb - \boldsymbol{q}_n ) \, F_n(\boldsymbol{q}_1,\dotsc, \boldsymbol{q}_n) \, \delta_{1} (\boldsymbol{q}_1) \dots \delta_{1} (\boldsymbol{q}_n) \,
 \\
& \theta_{n} (\boldsymbol{k})= \int_{q_1} \dotsi \int_{q_n} (2\pi)^3 \delta^3(\boldsymbol{k} - \boldsymbol{q}_1 - \dotsb - \boldsymbol{q}_n ) \, G_n(\boldsymbol{q}_1, \dotsc, \boldsymbol{q}_n) \, \delta_{1} (\boldsymbol{q}_1) \dots \delta_1 (\boldsymbol{q}_n) \, ,
\end{aligned} \end{equation}
where $\delta^3(\boldsymbol{k} -\boldsymbol{k'})$ denotes the three dimensional Dirac delta function.The superscript $3$ is used to differentiate it from the matter density function $\delta$. The SPT kernels can be calculated through the following recursion formulas
\begin{equation} \begin{aligned}
F_n(\q_1, \ldots ,\q_n) &= \sum_{m=1}^{n-1} { G_m(\q_1, \ldots ,\q_m)
 \over{(2n+3)(n-1)}} \Bigl[(2n+1) \alpha(\k_1,\k_2) F_{n-m}(\q_{m+1},
 \ldots ,\q_n) \nonumber \\ & +2 \beta(\k_1, \k_2)
 G_{n-m}(\q_{m+1}, \ldots ,\q_n) \Bigr] ,
\end{aligned} \end{equation}
\begin{equation} \begin{aligned}
G_n(\q_1, \ldots ,\q_n) &= \sum_{m=1}^{n-1} { G_m(\q_1, \ldots ,\q_m)
\over{(2n+3)(n-1)}} \Bigl[3 \alpha(\k_1,\k_2) F_{n-m}(\q_{m+1}, \ldots
,\q_n) \nonumber \\ &  +2n \beta(\k_1, \k_2) G_{n-m}(\q_{m+1},
\ldots ,\q_n) \Bigr].
\end{aligned} \end{equation}
where $\k_1=\q_1+ \ldots +\q_m$ and $\k_2=\q_{m+1}+ \ldots
+\q_n$; and $\alpha$ and $\beta$ are vertex functions associated with the non-linear terms in the coupled equations governing the fluid dynamics
\begin{equation}
\alpha(\k_1, \k_2) \equiv {\k_{12} \cdot \k_1
\over{ k_1^2}}, \ \ \ \ \ \beta(\k_1, \k_2) \equiv
{k_{12}^2 (\k_1 \cdot \k_2 )\over{2 k_1^2 k_2^2}}\, ,
\label{alpha-beta:eq}
\end{equation}
and we have defined, $\k_{12} \equiv \k_1 + \k_2$. In particular, the first non-trivial kernels are
\begin{equation} \begin{aligned}
F_2(\q_1,\q_2) &= \frac{5}{7} + \frac{1}{2} \frac{\q_1 \cdot
 \q_2}{q_1 q_2} (\frac{q_1}{q_2} + \frac{q_2}{q_1}) + \frac{2}{7}
 \frac{(\q_1 \cdot \q_2)^2}{q_1^2 q_2^2},
\\
G_2(\q_1,\q_2) &= \frac{3}{7} + \frac{1}{2} \frac{\q_1 \cdot
\q_2}{q_1 q_2} (\frac{q_1}{q_2} + \frac{q_2}{q_1}) + \frac{4}{7}
\frac{(\q_1 \cdot \q_2)^2}{q_1^2 q_2^2}.
\end{aligned} \end{equation}

The perturbation theory can be organized into the Feynman diagrams. For this purpose, as is customary, we depict these relations as

%\nolinenumbers
\begin{equation*}
\delta_n(\k)=
\raisebox{-55 pt}{
\centering
\begin{tikzpicture}[scale=.7]
\begin{scope}[>=Latex,line width=1.2 pt, dashed]
\draw[-](0,0) -- (2.8,2.8);
\draw[-](6,0) -- (3.2,2.8);
\draw[-](2,0) -- (2.9,2.8);
\end{scope}
\begin{scope}[>=Latex,line width=1 pt]
\draw[->](3,3.2) -- (3,4.2);
\draw[-](3,3.74) -- (3,4.75);
\draw (2.8,2.8)--(3.2,2.8)--(3.2,3.2)--(2.8,3.2)--(2.8,2.8);
\end{scope}

\draw (0,-.5) node  {$\q_1$};
\draw (2,-.5) node  {$\q_2$};
\draw (4,-.5) node  {$...\, \q_i\, ...$};
\draw (3.6,1.25) node  {$...$};
\draw (6,-.5) node  {$\q_n$};
\draw (3,5) node  {$\K$};

\end{tikzpicture}
}
\end{equation*}
%\linenumbers
In this diagram, the solid line denotes $\delta_n(\k)$, while every dashed line represents a linear density perturbation $\delta_{1}(\q)$. The vertices are indicated by the appropriate kernels, which can either be $F_{n}$ or $G_{n}$.

However, there is an alternative way of organizing the perturbation theory, which is sometimes more elucidating. Following \cite{Crocce_2005xy,bernardeau2013oda}, we introduce a doublet field
\begin{equation} \begin{aligned}
\Psi_a  = \left(\delta,-\dfrac{1}{\cal H} \theta \right)\,.
\end{aligned} \end{equation}
The equations of motion are
\begin{equation} \begin{aligned}
\label{main-eq}
\dfrac{\partial}{\partial \eta} \Psi_a (\bsb{k}, \eta)+ \Omega_{ab} (\eta) \Psi_b(\bsb{k}, \eta) =\int \frac{d^3 \k_1}{(2\pi)^3} \frac{d^3 \k_2}{(2\pi)^3}
\,\gamma^{(s)}_{abc}(\K,\K_1,\K_2)\,\Psi_b(\bsb{k}_1, \eta) \Psi_c(\bsb{k}_2, \eta),
\end{aligned} \end{equation}
in which, $\eta = \log a$. Specifically, for an Einstein-de Sitter Universe where $\Omega_m=1$, the following applies 
\begin{equation} \begin{aligned}
\Omega_{a\,b} =
 \Bigg[  \begin{array}{ccc}
0 & -1\\
-3/2 & 1/2
\end{array}
\Bigg]\,.
\end{aligned} \end{equation}
The non-vanishing components of the symmetrized vertex functions $\gamma^{(s)}$ are
\begin{equation} \begin{aligned}
\label{vert}
\gamma^{(s)}_{121} (\bsb{k},\bsb{k_1},\bsb{k_2}) &= \delta^3 (\bsb{k}-\bsb{k_1}-\bsb{k_2}) \,\alpha(\bsb{k_1},\bsb{k_2})/2\,,
\\
\gamma^{(s)}_{112} (\bsb{k},\bsb{k_1},\bsb{k_2}) &= \delta^3 (\bsb{k}-\bsb{k_1}-\bsb{k_2}) \,\alpha(\bsb{k_2},\bsb{k_1})/2\,,
\\
\gamma^{(s)}_{222} (\bsb{k},\bsb{k_1},\bsb{k_2}) &= \delta^3 (\bsb{k}-\bsb{k_1}-\bsb{k_2}) \beta(\bsb{k_1},\bsb{k_2})\,.
\end{aligned} \end{equation}
for the $\alpha$ and $\beta$ defined in Eq. \eqref{alpha-beta:eq}.A couple of field perturbations, lower-order in perturbation theory, can be mixed via the above vertices to build up a higher-order one.

\section{Fuzzy Dark Matter PERTURBATION THEORY} \label{sec:FDM}

Let us consider the following action for a real scalar field minimally coupled to the metric with canonical kinetic term and without self-interaction, as below \citep[see,][for a discussion]{witten}:
\begin{equation} \begin{aligned}
S = \int \frac{d^4x}{\hbar c^2}\sqrt{-g}\left [ \frac{1}{2} g^{\mu\nu} \partial_\mu \phi\partial_\nu \phi - \frac{1}{2}\frac{m_{\mathrm{f}}^2c^2}{\hbar^2}\phi^2 \right] \,.
\end{aligned} \end{equation}
Coherent oscillations of this field around the minimum of its potential will play the role of the dark matter in the universe, where the $m_{\mathrm{f}}$ is the mass of the FDM particles. In the non-relativistic limit, one can express $\phi$ in terms of a complex field $\psi$

\begin{equation} \begin{aligned}
\phi = \sqrt{\frac{\hbar ^3 c}{2m_{\mathrm{f}}}} \left ( \psi ^ \ast e^{-im_{\mathrm{f}}c^2t/\hbar} +  \psi e^{+im_{\mathrm{f}}c^2t/\hbar} \right ).
\end{aligned} \end{equation}

Now we substitute this definition into the Klein-Gordon equation for $\phi$ and use the perturbed Friedmann-Robertson-Walker metric

\begin{equation} \begin{aligned}
ds^2 = \left (1+ \frac{2\Phi}{c^2} \right ) c^2 dt^2 - a^2(t)\left ( 1- \frac{2\Phi}{c^2}\right ) d\mathbf{r}^2 \, ,
\end{aligned} \end{equation}

to arrive at the Schrodinger equation in an expanding universe

\begin{equation} \begin{aligned}
i\hbar\left(\dot\psi +\frac{3}{2}H\psi\right)&=\left(-\frac{\hbar^2}{2m_{\mathrm{f}}\,a^2}\nabla^2 +m_{\mathrm{f}}\Phi\right)\psi\,.
\end{aligned} \end{equation}

Note that in finding the above equation, considering the non-relativistic limit, we assumed \(\dot{\psi} \ll m_{\mathrm{f}}c^2 |\psi|/\hbar\) and \(\ddot{\psi} \ll m_{\mathrm{f}}c^2 |\dot{\psi}|/\hbar \). Here, \(a\) is the scale factor, $H$ is the Hubble parameter, and \(\Phi\) is the gravitational potential satisfying the Poisson's equation

\begin{equation} \begin{aligned}
\nabla^2\Phi=4\pi G a^2 \left (\rho - \Bar{\rho} \right ),
\label{poisson}
\end{aligned} \end{equation}

where \(\rho\) is the energy density of the scalar field, which in the non-relativistic limit is related to \(\psi\) by

\begin{equation} \begin{aligned}
\rho = m_{\mathrm{f}} |\psi|^2,
\end{aligned} \end{equation}
and $\Bar{\rho}$ is its mean value. Hence, the Schrodinger equation combined with Poisson's equation ultimately determines the dynamics of FDM in the non-relativistic limit, called the \textit{wave formulation} of the FDM dynamics.

Sometimes it is convenient to use another formulation for describing the FDM dynamics, namely \textit{the fluid formulation}; for example, when one is interested in the perturbation theory of FDM \citep{Hui}. For this purpose, one can use the so-called Madelung transformations,

\begin{equation} \begin{aligned}
\psi \equiv \sqrt{\frac{\rho}{m_{\mathrm{f}}} } e^{i\Theta} \quad , \quad \v \equiv \frac{\hbar}{m_{\mathrm{f}}\,a}\nabla \Theta = \frac{\hbar}{2iam_{\mathrm{f}}} \left ( \frac{\nabla \psi}{\psi} - \frac{\nabla \psi ^\ast}{\psi^\ast} \right ).
\label{madelung}
\end{aligned} \end{equation}

By the above substitution, the imaginary and real parts of the Schrodinger equation take the form of continuity and Euler equations, respectively,

\begin{equation} \begin{aligned}
\dot \rho + 3H\rho +\frac{1}{a}\nabla \cdot (\rho \boldsymbol{v} ) &=0 \, 
\label{cont:eq}
\end{aligned} \end{equation}
\begin{equation} \begin{aligned}
\Dot{\v} +H\v + \frac{1}{a} ( \v \cdot \nabla ) \v & = - \frac{1}{a} \nabla \Phi - \frac {\hbar ^2}{2m_{\mathrm{f}}^2\,a^3} \nabla p_Q , \label{euler:eq}
\end{aligned} \end{equation}
where $p_Q$ is the so-called quantum pressure that leads to the suppression of small-scale structures, given by

\begin{equation} \begin{aligned}
p_Q = - \frac{\nabla^2 \sqrt{\rho}}{\sqrt{\rho}}.
\end{aligned} \end{equation}

The continuity and Euler's equations, \eqref{cont:eq} - \eqref{euler:eq}, together with Poisson's equation,\eqref{poisson}, determine the dynamics of FDM in the fluid formulation. However, it should be noted that this formulation breaks down in the regions where FDM multi-streams occur because in these regions, the \textit{single} velocity in Eq. \eqref{madelung} is no longer well-defined \citep{uvcompletion,Vlasov}.

Now, we rewrite the continuity equation \eqref{cont:eq} and Euler's equation \eqref{euler:eq} in the Fourier space in terms of 
$\delta\equiv(\rho-\bar{\rho})/\bar{\rho}$ and $\theta\equiv\nabla.\boldsymbol{v}$. Here $\bar{\rho}$ denotes the average matter density, and $\delta$ is called local density contrast. The continuity equation is found as
\begin{align}
\label{eq:cont_full}
 \delta' (\bsb{k}, \eta)+ \theta(\bsb{k}, \eta)  =\int \frac{d^3 \k_1}{(2\pi)^3} \frac{d^3 \k_2}{(2\pi)^3}
\,\delta^3(\k-\k_1-\k_2)\,\alpha(\K_1,\K_2)\,\theta(\bsb{k}_1, \eta) \delta(\bsb{k}_2, \eta),
\end{align} 
and Euler's equation turns into
\begin{align}
\nonumber
\theta' (\bsb{k}, \eta)+& {\cal H}(\eta)\theta(\bsb{k}, \eta)+\dfrac{3}{2} \Omega_{m}(\eta) {\cal H}^2(\eta) \delta(\bsb{k}, \eta)  =
\\
&\int \frac{d^3 \k_1}{(2\pi)^3} \frac{d^3 \k_2}{(2\pi)^3}\,\delta^3(\k-\k_1-\k_2)
\,\beta(\K_1,\K_2)\,\theta(\bsb{k}_1, \eta) \theta(\bsb{k}_2, \eta)
+\dfrac{\hbar^2}{2 m_{\mathrm{f}}^2 a^2}\left[\nabla^2 \left( \frac{\nabla^2 \sqrt{1+\delta}}{\sqrt{1+\delta}} \right)\right]_{\K}.
\label{eq:Euler_full}
\end{align}
in which we have used Poisson's equation $\nabla^2\Phi=\frac{3}{2} \Omega_m {\cal H}^2\, \delta$ \citep{BernardeauSPT}. To simplify the QP term, note that for any given function $f$, the following identity holds true
\begin{equation} \begin{aligned}
    \nabla^2 \left( \frac{\nabla^2 f}{f} \right) = \dfrac{\nabla^2 (\nabla^2 f)}{f} -\dfrac{2 \nabla f.\nabla (\nabla^2 f)}{f^2} -\dfrac{(\nabla^2 f)(\nabla^2 f)}{f^2}+\dfrac{2 |\nabla f|^2 \nabla^2 f}{f^3}
\end{aligned} \end{equation}
Now, using the above equation for $f=\sqrt{1+\delta}$, we get
\begin{equation} \begin{aligned}
    \nabla^2 \left( \frac{\nabla^2 \sqrt{1+\delta}}{\sqrt{1+\delta}} \right) = \dfrac{1}{2}\,\nabla^2 \nabla^2 \delta-\dfrac{1}{2}\delta \,\nabla^2 \nabla^2 \delta-\dfrac{1}{2}(\nabla^2 \delta)(\nabla^2 \delta)-\dfrac{3}{2}\,(\partial_i \delta) \,\partial^i (\nabla^2 \delta) -\dfrac{1}{2} (\partial_i \partial_j \delta)(\partial^i \partial^j \delta) +...,
\end{aligned} \end{equation}
with summation over repeated indices understood, and we have $\nabla^2\equiv \partial_i \partial^i$. A comment is necessary. One may worry that the fourth-order derivatives in the QP term could violate the perturbative expansion for sufficiently small perturbations. However, as we will discuss at the end of this section, this term is proportional to $(k/k_J)^4$, so if we cut the theory on the scales much larger than Jean's wavelength, this factor would be much less than one. Again, the perturbation theory organized in the doublet representation reads as
\begin{equation} \begin{aligned}
\label{main-eq}
\dfrac{\partial}{\partial \eta} \Psi_a (\bsb{k}, \eta)+ \Omega_{ab} (\eta) \Psi_b(\bsb{k}, \eta) =\int \frac{d^3 \k_1}{(2\pi)^3} \frac{d^3 \k_2}{(2\pi)^3}
\,\gamma^{(s)}_{abc}(\K,\K_1,\K_2)\,\Psi_b(\bsb{k}_1, \eta) \Psi_c(\bsb{k}_2, \eta),
\\
+\delta_{a\,2}\,\sum_{n=1}^{\infty}\int \left(\prod_{i=1}^{n} \frac{d^3 \k_i}{(2\pi)^3} \right)
\,\Gamma^{(s)}_{n}(\K,\K_1,..,\K_n)\,\left(\prod_{i=1}^{n}\Psi_1(\bsb{k}_i, \eta)\right) 
\end{aligned} \end{equation}
where $\delta_{a\,2}$ in the second line above denotes the Kronecker delta, which is non-vanishing only for $a=2$. On the other hand, $\Omega_{a\,b}$ is slightly different for the FDM universe with an arbitrary $\Omega_m$, primarily due to the inclusion of a new term that accounts for QP
\begin{equation} \begin{aligned}
\Omega_{a\,b} =
 \Bigg[  \begin{array}{ccc}
0 & -1\\
-\dfrac{3}{2}\Omega_{m}(\eta)+\dfrac{\hbar^2\,k^4}{2m_{\mathrm{f}}^2 a^2(\eta) {\cal H}^2(\eta)} & 1+\dfrac{{\cal H}'(\eta)}{{\cal H}(\eta)}
\end{array}
\Bigg]\,.
\end{aligned} \end{equation}
The non-vanishing components of the symmetrized vertex function $\gamma^{(s)}$ are the same as \eqref{vert}. However, the QP leads to infinite new vertices, $\Gamma_n(\K,\K_1,..\K_n)$, which combine $n$ density fields to an $n$-th order velocity field. In particular, $\Gamma_2$ can be read as \cite{Hui}
\begin{equation}
\begin{aligned}
    \Gamma_2(\K,\K_1,\K_2) = \dfrac{-\hbar^2}{8 m_{\mathrm{f}}^2 a^2 \mathcal{H}^2} \left[ (k_{1}^2+k_2^2)^2+3\,(\K_1.\K_2)(k_1^2+k_2^2)+2\,(\K_1.\K_2)^2\right]\,\delta^3 (\bsb{k}-\bsb{k_1}-\bsb{k_2}) 
\end{aligned}
\end{equation}

%After some algebra, we find that this new vertex modifies the $F_2$ by the following amount
%\begin{equation}
%\begin{aligned}
%    \Tilde{F}_2(\K_1,\K_2) = \dfrac{-\hbar^2}{132 m_{\mathrm{f}}^2\,H^2} \left[ (k_{1}^2+k_2^2)^2+3\,(\K_1.\K_2)(k_1^2+k_2^2)+2\,(\K_1.\K_2)^2\right]
%\end{aligned}
%\end{equation}
 Let us analyze the behavior of the above vertex for the soft-- long wavelength-- outgoing momentum.  In the soft external momentum limit, $k_L=|\k_1+\k_2|\ll k_1,k_2$ we see that
\begin{align}
    \Gamma_2(\K_1,\K_2) \sim \dfrac{-\hbar^2 (k_S/a)^4}{8\, m_{\mathrm{f}}^2\,H^2} \left(\dfrac{k_L^2}{k_S^2} \right) 
\end{align}
where $k_2\simeq k_1\equiv k_S$. The above result shows that the new non-trivial UV physics of FDM complies with the ``double softness" rule. The double softness means that short wavelength fluctuations, with wavenumber $k_S$, can mix to form a longer wavelength perturbation, with momentum $k_{\mathrm{L}}$, that is suppressed at least by $k_{\mathrm{L}}^2$. Note that the double softness results from local interaction and momentum conservation, making it a crucial guideline for writing the EFT expansion.
One might worry that the argument above fails since the factor by which $(k_L/k_S)^2$ is multiplied can grow arbitrarily for sufficiently large $k_S$ values. Additionally, aside from an overall numerical factor, the same term appears in $\Omega_{ab}$-- that is, the ratio of the linearized QP term to the gravity force term in Euler's equation. Noting that (comoving) Jeans momentum is defined as $k_J \equiv \ a \sqrt{6^{1/2}\,H m_{\mathrm{f}}/\hbar} $, this factor can be written as
\begin{align}
\dfrac{\hbar^2 (k_S/a)^4}{8\, m_{\mathrm{f}}^2\,H^2}= \dfrac{3}{2} \left(\dfrac{k_S}{k_J}\right)^4
\end{align}
Hence, if we cut the theory for momenta much smaller than the (comoving) Jeans momentum $k_J \propto a \sqrt{H m_{\mathrm{f}}/\hbar}$ within the EFT approach, these two corrections vanish. Note that, at $z=0$, the Jeans wavenumber is much larger than the nonlinear wavenumber $k_J\gg k_{NL}$ for viable FDM mass values\footnote{One could argue that this model has a natural cut-off, as the power spectrum is suppressed for wavelengths exceeding Jean's wavelength. However, as discussed, a viable perturbation theory requires cutting the theory at scales much below the non-linear scale $k_{\mathrm{NL}}$.}.
One may worry about whether this inequality holds at early times since comoving Jean's wavenumber goes as $a^{1/4}$ \citep[see, e.g.][]{Hui}. Recalling that $k_J(t_{\mathrm{eq.}})\simeq 9 \times (m_{\mathrm{f}}/10^{-22}\, \mathrm{eV})^{1/2}\,\mathrm{Mpc}^{-1}$, even at the equality and for boson as light as $m_{\mathrm{f}}=10^{-23}\,\mathrm{eV}$ the Jeans wavenumber is much larger $k_{\mathrm{NL}}$.  Consequently, SPT kernels can be used for perturbation theory in the FDM universe at scales larger than $k_{\mathrm{NL}}^{-1}$.

\subsection{Linear Perturbation Theory}
\label{sec:lin_perturb_theory}
On scales longer than the non-linear scale, the density contrast $\delta\equiv(\rho- \Bar{\rho})/\Bar{\rho}$, the peculiar velocity $v$ and $\Phi$ are small. At the first order of the perturbation theory, Eqs. \eqref{eq:cont_full} and \eqref{eq:Euler_full} can be read as
\begin{equation} \begin{aligned}
\delta'+ \theta &=0 \, 
\end{aligned} \end{equation}
\begin{equation} \begin{aligned}
\theta' +{\cal H} \theta \ + \dfrac{3}{2} \Omega_m {\cal H}^2 \delta & =  \dfrac{\hbar^2}{4 m_{\mathrm{f}}^2\, a^2} \nabla^2 \nabla^2 \delta  \label{fuzzy-linear:eq}
\end{aligned} \end{equation}
The coupled equations above are the same as the equations governing the dynamics of CDM perturbations except for a pressure-like term in the Euler equation.

The matter density power spectrum is defined as
\begin{equation}
    P(k;z)=\frac{1}{(2\pi)^3} \int d^3\boldsymbol{r}\, \xi(\boldsymbol{r};z) \,e^{-i\boldsymbol{k}\cdot \mathbf{r}}  ,
\end{equation}
in which, $\xi(\boldsymbol{r};z)$ represents the two-point correlation function of the density contrast at redshift $z$ 
\begin{equation}
    \xi(\boldsymbol{r};z)= \langle \delta(\boldsymbol{x};z) \delta(\boldsymbol{x}+\boldsymbol{r};z)\rangle\, .
\end{equation}
Or equivalently, we have 
\begin{align}
     \langle \delta(\k;z) \delta(\k';z)\rangle=P(k;z) \delta^3(\k-\k')
\end{align}
where $\delta(\k;z)$ is the Fourier transform of matter density contrast $\delta(\x;z)$. We calculate the power spectrum by plugging in the series expansion for matter density as
\begin{align}
P(k;z)= P_{11}(k;z)+P_{22}(k;z)+P_{13}(k;z)+...,
\end{align}
where $P_{mn}(k;z)\equiv \langle \delta_m(k;z)\delta_n(k;z) \rangle$, and $\delta_{m}$ is the $m-$order term in the perturbation expansion \eqref{eq:perturb-exp}. Note that $P_{mn}$ will only have a non-zero value when the sum of $m$ and $n$ is an even number. \citep{Bernardeau:2001qr}.

At the linear order, the suppression of the FDM power spectrum relative to the CDM - due to the QP - can be characterized by a transfer function, shown below \citep{Hu_2000}
\begin{equation} \begin{aligned}\label{linear-power-fuzzy}
P_{\mathrm{FDM}}(k,z)=\left[\frac{\mathcal{T}_{\mathrm FDM}(k,z)}{\mathcal{T}_{\mathrm CDM}(k,z)}\right]^2 P_{\mathrm{CDM}}(k,z) = \mathcal{T}^2(k,z)P_{\mathrm{CDM}}(k,z)\, ,
\end{aligned} \end{equation}
in which $P_{\mathrm{FDM}}(k,z)$ and $P_{\mathrm{CDM}}(k,z)$ are the three-dimensional matter power spectra of FDM and CDM, respectively, at the redshift $z$, and the comoving wave number is represented by $k$. The transfer function $ \mathcal{T}(k,z) $ can be well approximated by the redshift-independent expression. In other words, the transfer function can be factorized into a growth function depending only on time and a time-independent transfer function $\cal T$.
\begin{equation} \begin{aligned} \label{power2}
\mathcal{T}(k)=\frac{\cos{x^3}}{1+x^8},\quad \textrm{where:} \quad x=1.61 \times \left(\frac{m_{\mathrm{f}}}{10^{-22}\,\mathrm{\mathrm{eV}}}\right)^{1/18} \times \frac{k}{k_J(t_{\mathrm{eq.}})}\, .
\end{aligned} \end{equation}
in which the parameter $k_J(t_{\mathrm{eq.}})\simeq9 \times (m_{\mathrm{f}}/10^{-22}\, \mathrm{eV})^{1/2}\,\mathrm{Mpc}^{-1}$ is the critical scale of Jeans wavenumber at the matter-radiation equality.

\begin{figure*}[t]
\gridline{\fig{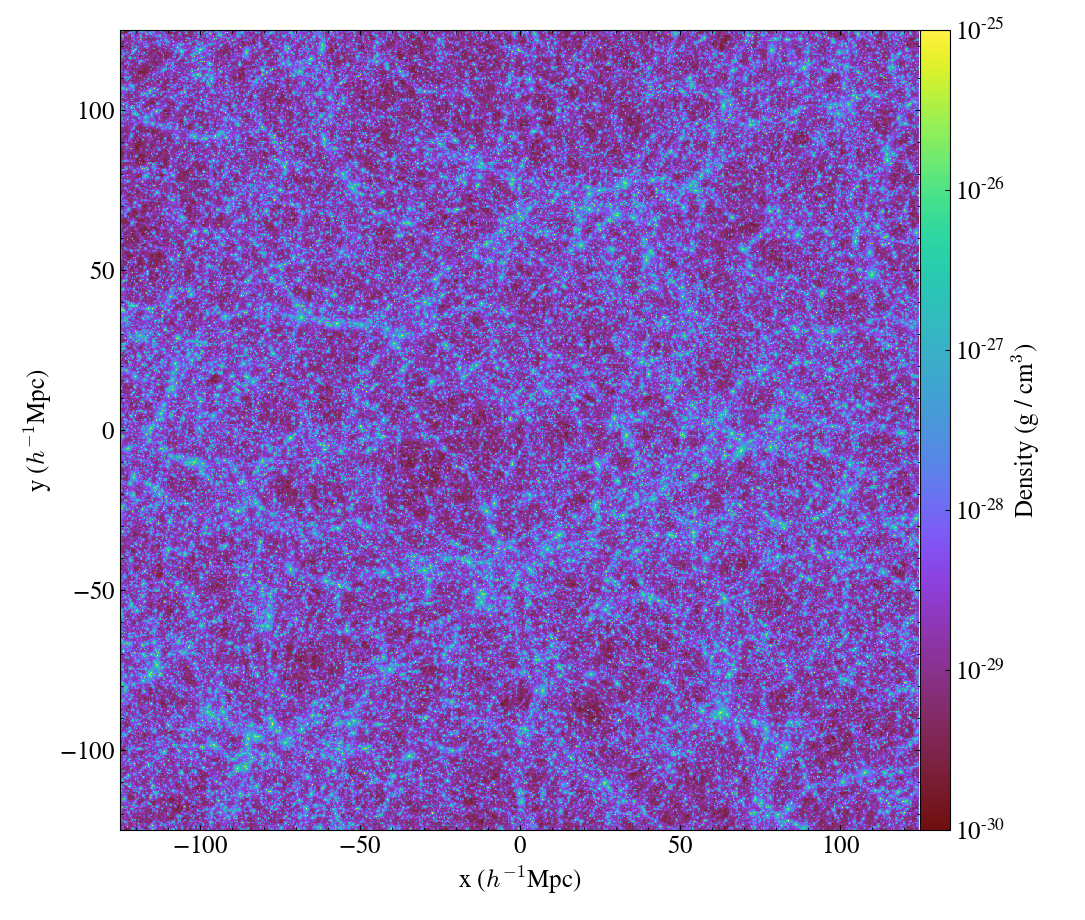}{0.48\textwidth}{(a)}
          \fig{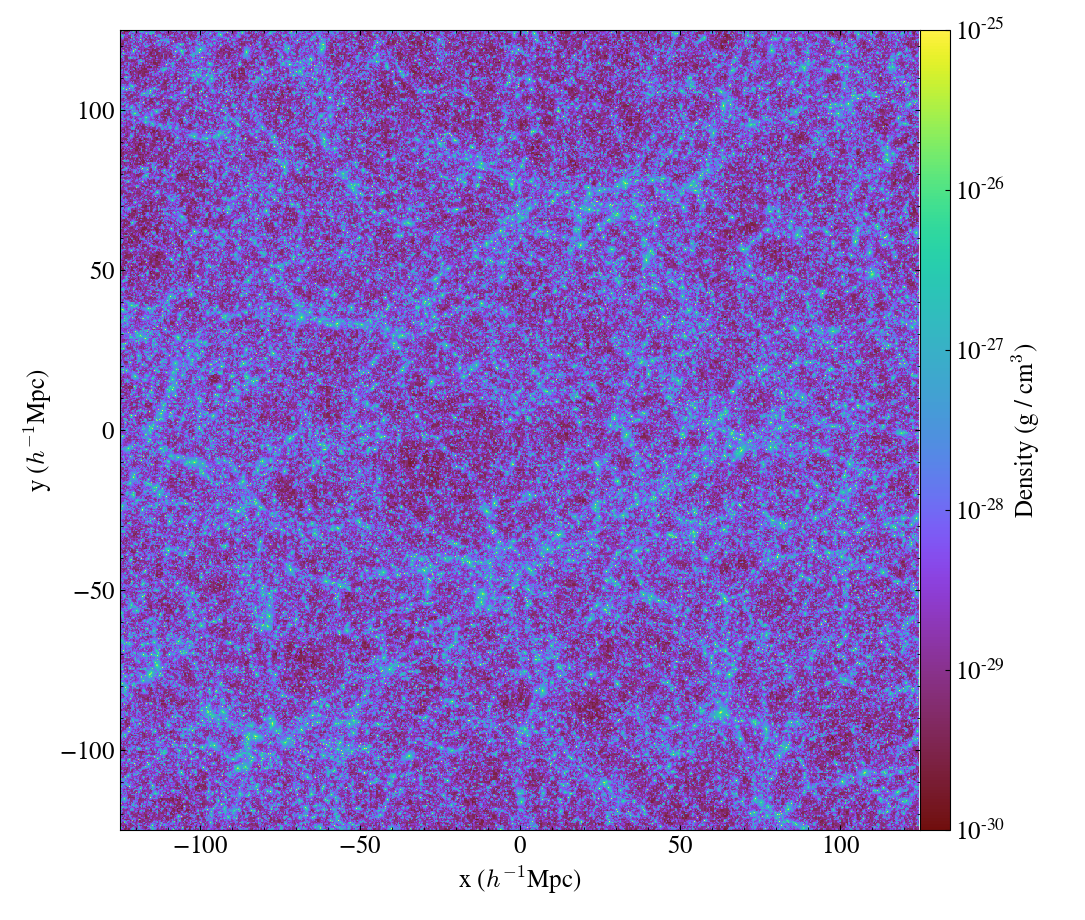}{0.48\textwidth}{(b)}
          }
\caption{Projected density plots of (a) CDM and (b) FDM ($m_{22}=0.1$) simulations, at $z=0$, were performed using Gadget-2 code with the proper initial conditions for each of them. They both have $512^3$ particles and $250\,h^{-1}\,\mathrm{Mpc}$ length of box. To start with the \textit{same} realizations, we use the same random seed number to generate the initial conditions. Consequently, only tiny visual differences can be found between the figures, resulting from the suppressed FDM transfer function. However, EFT could systematically parameterize these tiny differences on large scales.}
\label{fig:250plots}
\end{figure*}

\section{The Cosmological Simulations} \label{sec:sim}

We compare the predictions for the matter power spectrum from the EFT of LSS in the cases of CDM and FDM. In particular, we use cosmological simulations to determine the EFT parameters of CDM and FDM in 1-loop order. These simulations must be performed on a box large enough to encompass the quasi-linear regime, namely $k \sim 0.1\,h\,\mathrm{Mpc}^{-1}$. Recently, several FDM cosmological simulation codes have been developed that use different approaches to follow the FDM dynamics \citep[see,][]{FDMsimulations}. A primary class of these simulations solves the Schrodinger-Poisson equations for an expanding universe. Several papers use wave formulation to perform cosmological simulations \citep[e.g.][]{interference,Hui,axionyx,Simon}. In this sort of simulation, the small-scale fingerprints of the FDM, such as the solitonic cores and interference patterns, are well captured. However, these simulations fail to study the larger scales. This limitation arises from the fact that the velocity, as defined in Eq. \eqref{madelung}, is determined by the gradient of the wave function's phase. Consequently, the velocity cannot surpass a maximum value due to the restriction imposed by the maximum phase difference of $2\pi$ between neighboring grids in the simulation. Consequently, when employing the wave-based approach of the FDM in simulations, the grid sizes must not exceed the de-Broglie wavelength \citep{Hui,Simon}. As a result, conducting simulations involving large boxes necessitates increasingly substantial computational resources. Today, the largest FDM simulations ever performed using the wave formulation have a box size of the order of $\sim 10\,h^{-1}\,\mathrm{Mpc}$ and are reliable only down to $z \sim 3$ \citep[see, e.g.][]{Simon}. Therefore, the reachable box size of this sort of simulation is not yet sufficient to study quasi-linear scales, i.e.$\gtrsim 200\,h^{-1}\,\mathrm{Mpc}$.

\begin{figure}[t]
\centering
\scalebox{0.8}{\plotone{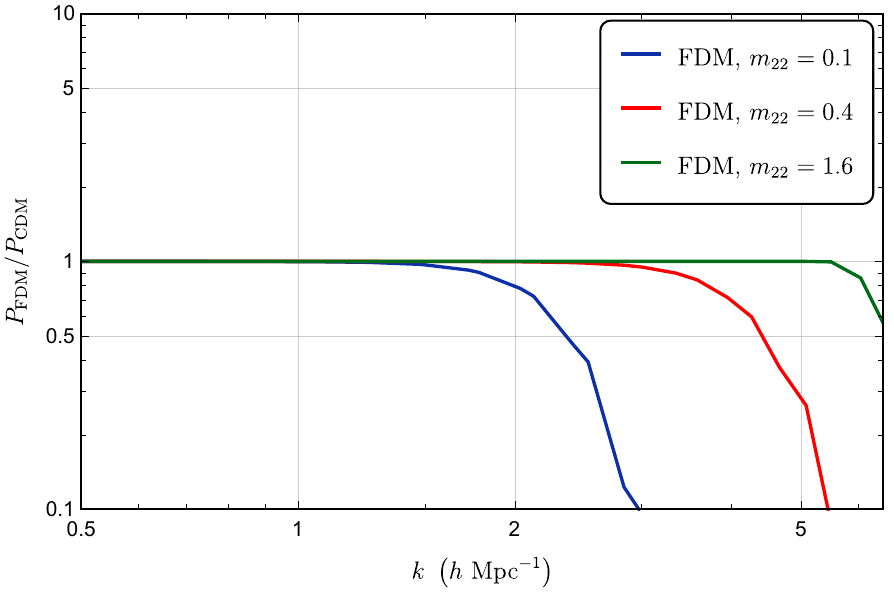}}
\caption{The power spectra of FDM Simulations with different masses at the initial redshift, i.e., $z=99$, normalized by CDM. The wave number in which FDM deviates from CDM is smaller for lower FDM masses.} 
\label{fig:power250}
\end{figure}

The second approach is to use the fluid formulation of the FDM dynamics. In this approach, the Smoothed-Particle-Hydrodynamics (SPH) methods \cite{particleincell,mocz,ax-gadget} are used to calculate the extra force due to the gradient of the QP term in Eq.\eqref{euler:eq} on the FDM particles in an N-body simulation. Although this approach could not reproduce the interference patterns and has some intrinsic inaccuracies in small scales \citep[see, e.g.,][]{FDMsimulations},  it is still suitable for studying large-scale structure formation. For instance, simulations with a box size of $50\,h^{-1}\,\mathrm{Mpc}$ are performed in \citep[see][]{zhanglym}. However, generating reliable larger simulations that could encompass quasi-linear scales is still beyond the capabilities of these codes \citep{FDMsimulations}.

Another alternative is to use the FDM initial conditions for the ordinary CDM cosmological simulation codes.  This has been shown to be a good approximation if we are interested in large-scale structure formation. The difference between the mass power spectra of a full FDM simulation and a simulation with only FDM initial conditions is well below the percent level \citep[see, e.g.][]{ax-gadget}. For the FDM  of mass of orders $\sim 10^{-23}-10^{-22}\, \mathrm{eV}$, for wavenumbers $k\sim 1-10\,h\,\mathrm{Mpc}^{-1}$ and smaller, the difference becomes utterly negligible. This fact, together with the ability of CDM codes to successfully simulate quasi-linear scales, makes this approach suitable for our present purpose.

We performed simulations using the publicly available \emph{Gadget-2} code. The \emph{N-GenIC} generates the initial conditions  \footnote{\url{ https://gitlab.Mpcdf.mpg.de/rwein/ngenic} }, where we have also implemented the Eq. \eqref{power2} to generate the suppressed FDM initial conditions\footnote{
While there are more accurate methods available, such as the axionCAMB \citep{CAMB} and Class.FreeSF \citep{ClassFSC} codes, our method still holds its own. On the scale where the suppression occurs in the matter power spectrum of the FDM, there are minor discrepancies between the predictions of these codes and the method we have used, and even between these codes themselves\citep[e.g.,][]{suppression,Urena-Lopez:2019kud}. However, these minor discrepancies in the shape and slope of the suppression at the scale $k \sim 5h/\mathrm{Mpc}$ would have a negligible impact on our analysis in the quasi-nonlinear scales, at least at the level of the precision of our simulations.}. The cosmological parameters used in the simulations are ${\{\Omega_m,\Omega_b, \Omega_\Lambda,h,n_s,\sigma_8\}=\{0.295,0.0468,0.705,0.688,0.9676,0.835\}}$. the simulations have $512^3$ particles and a box size of $250\,h^{-1}\,\mathrm{Mpc}$. We performed three FDM simulations with three different masses, namely $m_{22}= 0.1,\,0.4\,\mathrm{and}\,1.6$, where $m_{22}$ is defined as $m_{22} \equiv m_{\mathrm{f}}/10^{-22}\, \mathrm{eV}$. One needs about 28,000 core hours of computing resources to perform these simulations. For this purpose, we used the Sharif University of Technology's High-Performance Computing Center (HPC) machine.

In Fig. \eqref{fig:250plots}, we compare the slice-projection plots of the CDM and FDM simulations with the smallest mass at $z=0$. Due to the same random seed number used to generate the initial conditions, the plots appear superficially similar, ensuring that we started with the \textit{same} realizations. Nevertheless, the tiny visual differences, that could be encoded in the speed of sound parameters, are rooted in the FDM initial power spectrum mixed up in the subsequent non-linear dynamics. 

Fig. \eqref{fig:power250} shows the matter power spectra of the FDM simulations at the initial redshift, i.e., $z=99$. As expected, the FDM power spectrum deviates from the CDM in larger scales for smaller masses. 

\section{Power Spectrum: Effective Field Theory}\label{sec:EFT}
We use the effective field theory approach to theoretically predict the large-scale perturbations of the large-scale structure of the Universe. We smooth out perturbations on a scale of $\Lambda^{-1}$, which is equivalent to including perturbations with momentum less than a UV scale of $\Lambda$ in perturbation theory. The smallness of the \emph{smoothed} density and velocity on large scales suggests that appropriate perturbation theory should \emph{converge} to the correct answer in this regime. EFT of LSS provides us with theoretical predictions well beyond the validity range of SPT. However, this comes at the cost of introducing an infinite number of effective interactions into the fluid equations of Standard Perturbation Theory (SPT).
The conservation of momentum and the locality of the short-scale dynamics guarantee that the short fluctuations can, at most, affect longer-wavelength perturbations at $k^2$ order, regardless of whatever physics holds on the UV scale \cite{Abolhasani_2016}. That is crucial for the effective field theory to give a viable description at a large scale when either we do not precisely know the physics governing the UV scales or the UV physics is too complicated to be tracked. In this approach, the momentum integral in diagrams should be cut at scale $\Lambda$ so that higher-order field fluctuations would be cut-off dependent. However, the observed physical quantities do not depend on our chosen cut-off. These dependencies must be exactly canceled by the appropriate counterterms coming from integrating our UV physics- namely, where the SPT does not work- within the context of the effective field theory. In this article, we specifically examine the leading order correction in derivatives, which is the sound speed term. This term acts as a quantitative measure of how UV physics affects large-scale perturbations. It's worth noting that two models with the same effective sound speed may have different power spectra if their linear power spectra vary. Therefore, sound speed serves as a concise measure that captures all UV physics, in the leading order, within one number(For a more detailed discussion, See App. \ref{1:app})

\begin{figure}[t!]
	\centering
	\scalebox{0.85}{\plotone{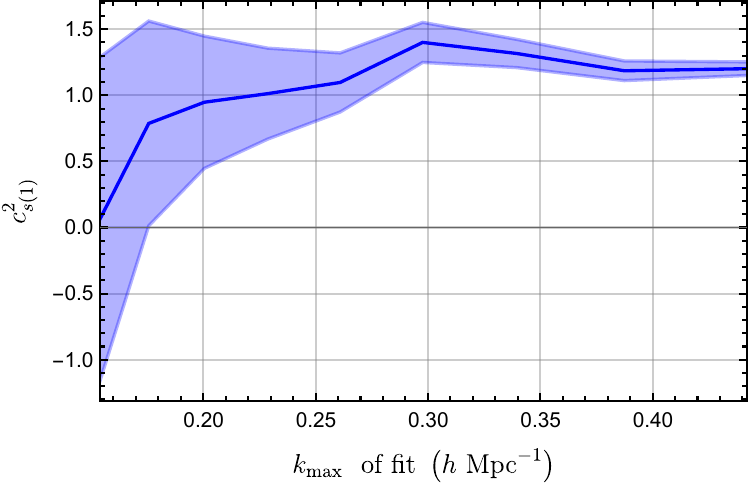}}
	\caption{The value of the one-loop speed of sound parameter obtained from our CDM simulation when using different upper bounds for the fitting interval ($k_{max}$). The shaded blue region represents the 2-$\sigma$ error bars. As discussed in the text, the value $k_{max}\simeq 0.30$ at which the $c_{s(1)}^2$ exits for the first time from the previous error bars, should be chosen as the appropriate upper bound ($k_{\mathrm{fit}}$)}
	\label{cs-kmaxfit}
\end{figure}

The prediction of EFT for the one-loop matter power spectrum, incorporating the contribution corresponding to the so-called speed of sound, is \citep{twoloops}
\begin{equation} \begin{aligned}
P_\text{EFT-1-loop}(k,z) = P_\text{SPT-1-loop}(k,z) + P^{(c_s)}_{\mathrm{tree}}(k,z)\ ,
\label{eq:peft1loop}
\end{aligned} \end{equation}
where
\begin{equation} \begin{aligned}
P_\text{SPT-1-loop}(k,z) = [D_1(z)]^2 P_{11}(k) + [D_1(z)]^4 P_{1-\mathrm{loop}}(k)\ ,
\end{aligned} \end{equation}
and
\begin{equation} \begin{aligned}
P^{(c_s)}_{\mathrm{tree}}(k,z) = -2(2\pi) c^2_{s(1)}(z) [D_1(z)]^2 \frac{k^2}{k_{\mathrm{NL}^2}} P_{11}(k)\ ,
\end{aligned} \end{equation}
$P_{\textmd{1-loop}}(k)$ is the one-loop correction to the linear power spectrum in SPT, $D_1(z)$ is the linear growth function defined as $\delta(k,z) = D_1(z) \delta(k,0)$, $k_{\textrm{NL}}$ is the non-linear scale at which $\delta(k,0)$ becomes of the order of unity, and $c_{s(1)}^2$ is the so-called one-loop speed of sound\footnote{Despite being referred to as "speed," the parameter does not possess the dimension of length per time. In fact, it is the \textit{dimensionless} speed of sound, obtained by multiplication of the factor $k^2/{{\cal H}_0}^2$ \citep[see,][for more details]{Hertzberg_2014}.} parameter that determines the magnitude of the counterterm introduced by EFT of LSS at the one-loop level. $P_{\textmd{1-loop}}(k)$ is given by \citep{irresummed}
\begin{equation}\label{p1loop}
P_{\textmd{1-loop}}(k) = P_{22} (k) + P_{13} (k) \, ,
\end{equation}
where
\begin{equation}\label{P22}
P_{22} (k) = \frac{k^3}{2 \pi^2} \int \dd r \,r^2 \int \dd x P_{11}(r k) P_{11}(k \sqrt{r^2 - 2 r x + 1}) \left( \frac{7 x + (3 - 10 x^2)r}{14 r (r^2 - 2 r x + 1)}\right)^2 \, ,
\end{equation}
and
\begin{equation}\label{P13}
\begin{aligned}
P_{13} (k) = \frac{k^3}{252 (2 \pi)^2} P_{11} (k) \int \dd r\, r^2 P_{11} (k r) \Bigg[ \frac{12}{r^4} &- \frac{158}{r^2} + 100 - 42 r^2 \\ &+ \frac{3}{r^5} \left(7 r^2 + 2\right) \left( r^2 - 1\right)^3 \ln \left|\frac{1+r}{1-r}\right|\Bigg] \, .
\end{aligned}
\end{equation}
We calculate the linear power spectrum, $P_{11}(k)$, using the \textsc{CAMB}\footnote{ \url{https://camb.info/} } code with the same cosmological parameters as of our simulations.\\

\subsection{Comparing the speed of sound for \rm{CDM} and \rm{FDM}}

\begin{figure}[t!]
	\centering
	\scalebox{0.75}{\plotone{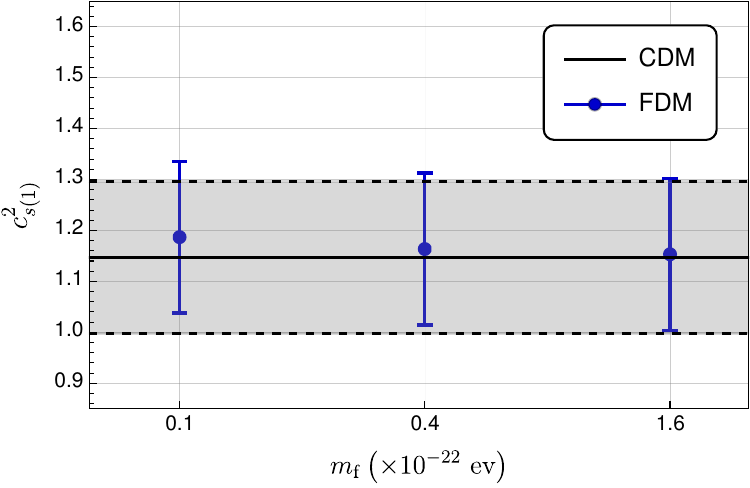}}
	\caption{The value of the sound speed for different FDM masses is shown here. The black line and the shaded area are the parameter values for the CDM simulation; 1-$\sigma$ intervals are shown with bars.}
	\label{fig:cs-kmaxfit}
\end{figure}

In order to determine the best-fit values of $c_{s(1)}^2$, we fit Eq. \eqref{eq:peft1loop} to the non-linear power spectrum obtained from either simulations or observations. However, the best "fitting interval" is not known a priori from the theory. \cite{precision} proposes a systematic way to find the appropriate maximum wavenumber of the fitting interval, namely $k_{\mathrm{fit}}$. In this procedure, by gradually increasing the maximum wavenumber of the fitting interval, $k_{max}$, we arrive at a wavenumber at which the best-fit value of $c_{s(1)}^2$ exceeds the error bars of the previous ones obtained for smaller $k_{max}$'s. We use this wave number as $k_{\mathrm{fit}}$. As shown in Fig. \eqref{cs-kmaxfit}, following this procedure using the power spectrum of our simulations, we arrive at the value $k_{\mathrm{fit}} = 0.28\,h\,\mathrm{Mpc}^{-1}$. So, we use this value as the upper bound of the fitting intervals in the following calculations.

We determine the value of $c_{s(1)}^2$ using our CDM and FDM N-body simulations and then compare them. The simulations' box size is $L= 250 \,h^{-1}\,\mathrm{Mpc}$ with $512^3$ particles. Since we perform finite box/finite simulation, only a finite number of modes are at hand, so to calculate the integrals in Eqs.\eqref{P22}-\eqref{P13}, we use the linear CAMB data for $P_{11}(k)$ as an approximation. 

\begin{center}
\begin{table}[ht!]
    \begin{tabular}{>{\centering\arraybackslash}m{2.9cm}|>{\centering\arraybackslash}m{1.9cm}|>{\centering}m{1.9cm}|>{\centering\arraybackslash}m{1.9cm}|>{\centering\arraybackslash}m{1.9cm}}

       Simulation \vspace{0.13cm} & CDM \vspace{0.13cm} & FDM ($m_{22}=1.6$) \vspace{0.03cm} & FDM ($m_{22}=0.4$) \vspace{0.03cm} & FDM ($m_{22}=0.1$) \vspace{0.03cm} \\
       \hline
        \vspace{0.15cm} $c_{s(1)}^2$ \vspace{0.15cm} & \vspace{0.15cm} 1.14 \textpm 0.15 \vspace{0.15cm} & \vspace{0.15cm} 1.15 \textpm 0.15 \vspace{0.15cm} & \vspace{0.15cm} 1.16 \textpm 0.15 \vspace{0.15cm} & \vspace{0.15cm} 1.18 \textpm 0.15 \vspace{0.15cm} \\
       \hline
        \vspace{0.05cm} Difference Relative to CDM \vspace{0.05cm} & - \vspace{0.15cm} & +1 \% \vspace{0.15cm} & +2 \%  \vspace{0.15cm} & +4 \% \vspace{0.15cm}\\

    \end{tabular}
    \caption{The value of $c_{s(1)}^2$ in the units of $ \left(k_{\mathrm{NL}}/(2\,h\,\mathrm{Mpc}^{-1}) \right)^2$ for the CDM and FDM simulations.}
    \label{tab:cs-compar-spt}
\end{table}
\end{center}

By choosing the fitting interval to be $k\in\left[2 \pi/L, 0.28\right]\,h\,\mathrm{Mpc}^{-1}$, we get the best-fit value for the effective sound speed for CDM simulation to be \footnote{In our analysis, we employ the NonlinearModelFit function available in \textit{Mathematica} to perform the fitting procedure and estimate confidence intervals. This function utilizes a nonlinear least-$\chi^2$ method to fit the model to the given data, the power spectrum of the simulations in our case. The resulting output provides the square roots of the diagonal elements of the covariance matrix, which estimate the standard deviation of the best-fit values of the model's parameters (the speed of sound parameter in our case.)}
\begin{equation} \begin{aligned}
c_{s(1)}^2 = 1.14 \pm 0.15 \left(k_{\mathrm{NL}}/(2\,h\,\mathrm{Mpc}^{-1}) \right)^2.  
\end{aligned} \end{equation}

This result agrees very well with that of the previous studies \citep[see, e.g. ][]{irresummed,twoloops,allredshifts}. For instance, \cite{allredshifts} found $c_{s(1)}^2 = 1.05^{+0.05}_{-0.27} \left(k_{\mathrm{NL}}/(2\,h\,\mathrm{Mpc}^{-1}) \right)^2$ for a universe with $\sigma_{8}=0.81$. If we use the scaling relation, $c_{s(1)}^2 \propto \sigma_{8}^{3.5}$, suggested in that paper, the speed of sound translates to $c_{s(1)}^2 = 1.17^{+0.06}_{-0.30} \left(k_{\mathrm{NL}}/(2\,h\,\mathrm{Mpc}^{-1}) \right)^2$ for a universe with the same $\sigma_{8}$ as ours, showing complete agreement between the results.

We now repeat the same procedure for our FDM simulations. The linear power spectrum, $P_{11}$, should be modified for the FDM case via Eq. \eqref{linear-power-fuzzy}. This is a good approximation for studying the scales we are interested in; as argued in section \ref{sec:lin_perturb_theory}, on large scales, one could use the CDM SPT kernels, Eqs. \eqref{P22}-\eqref{P13}, for the FDM.

A remark is necessary. The back-reaction from the suppressed small-scale modes may slightly reduce the power spectrum, not accounted for in the SPT terms. Since the EFT counterterm has a minus sign, suppressing the power results from a higher speed of sound. Therefore, QP contributes positively to the speed of sound.

Repeating the fitting procedure for the FDM simulation with a mass of $m_{22}=1.6$, the best-fit value for the speed of sound is found to be $c_{s(1)}^2 = 1.16 \pm 0.15\left(k_{\mathrm{NL}}/(2\,h\,\mathrm{Mpc}^{-1}) \right)^2$. This value doesn't show a significant deviation from the value obtained for the CDM simulation.

Using FDM simulations with masses of $m_{22}=0.4$ and  $m_{22}=0.1$, we get the values $c_{s(1)}^2 = 1.16 \pm 0.15\left(k_{\mathrm{NL}}/(2\,h\,\mathrm{Mpc}^{-1}) \right)^2$ and $c_{s(1)}^2 = 1.18 \pm 0.15\left(k_{\mathrm{NL}}/(2\,h\,\mathrm{Mpc}^{-1}) \right)^2$, respectively. The results are listed in Table \ref{tab:cs-compar-spt} and depicted in Fig.\eqref{fig:cs-kmaxfit}.

While our results show that there is no difference between the speed of sound values for CDM and FDM models, at least not beyond the error bars resulting from our relatively low-resolution simulations, there appears to be an increasing trend in the mean values of speed of sound as the FDM mass decreases.  \\

\begin{figure}[t!]
	\centering
	\scalebox{1.1}{\plotone{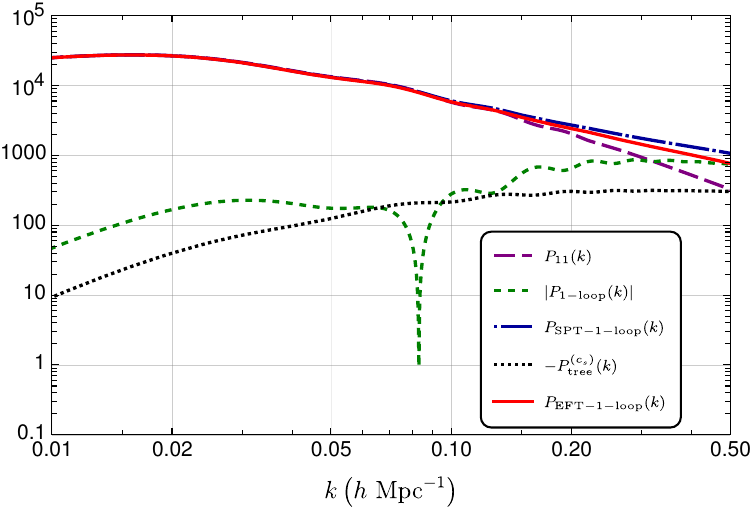}}
	\caption{One-loop power spectrum for FDM and the comparison of its different contributing terms. We use the ordinary CDM SPT kernels, as justified in the text, and the Eq. \eqref{linear-power-fuzzy} for the linear power spectrum, $P_{11}(k)$.}
	\label{fig:P22P13}
\end{figure}

\begin{figure}[hb!]

\begin{subfigure}[CDM]{\includegraphics[width= 3.5in]{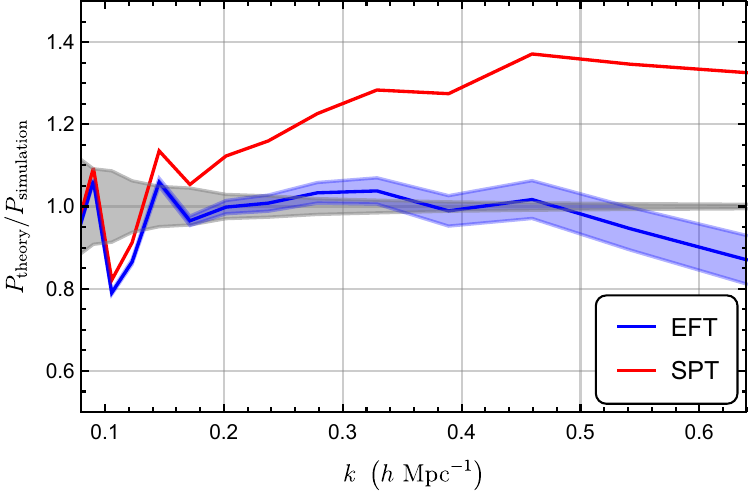}}
\label{fig:cdm}
\end{subfigure}
\begin{subfigure}[FDM ($m_{22}=0.1$)]{\includegraphics[width= 3.5in]{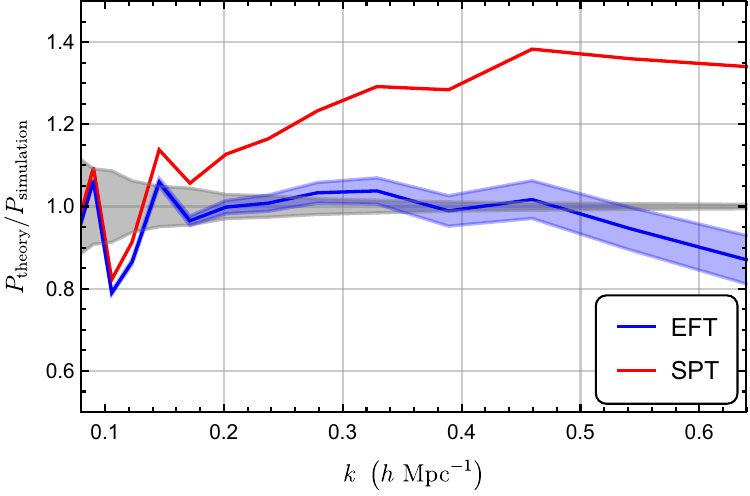}}
\label{fig:fdm1}
\end{subfigure}

\caption{EFT prediction for the matter power spectrum at one-loop order at $z=0$, for (a) the CDM simulation and (b) the FDM simulation with $m_{22}=0.1$. The grey-shaded regions depict the standard error of the simulations, and the EFT error bars are shaded in blue. The fractional difference of the SPT is also shown for comparison. As discussed in the text, the linear power spectrum used in the SPT and EFT formula is the CAMB linear data. The plots are superficially identical since the differences are at the percent level.}
\label{fig:frac-diff-Simulation-0.3}
\end{figure}

The FDM one-loop power spectrum is shown in Fig.\eqref{fig:P22P13}. The different contributions are plotted separately to compare their magnitude at different scales. One can see that the contribution of the one-loop corrections dominates over that of the linear term at the scales $k\approx 0.3\,h\,\mathrm{Mpc}^{-1}$. Fig.\eqref{fig:frac-diff-Simulation-0.3} compares the SPT and EFT predictions for the matter power spectrum, divided by the power calculated from the simulation. As expected, one can push the theory's validity toward the small scales using the EFT of LSS. While the SPT's power spectrum deviates from the simulation by more than 2-$\sigma$ at $k\approx 0.20\,h\,\mathrm{Mpc}^{-1}$, EFT predictions are consistent with the full non-linear simulation at 1-$\sigma$ even at $k\approx 0.54\,h\,\mathrm{Mpc}^{-1}$. We can extend the domain of applicability of the theory by considering higher loops or higher orders in perturbation theory. However, doing so comes with a cost of introducing new counter terms, which is equivalent to adding terms of order $(k/k_{\mathrm{NL}})^4$ and higher. On the other hand, for a constant $k$, going higher in the perturbation theory leads to smaller errors, as expected \citep{precision}.

\section{Summary and Discussion}

This paper compares the statistics of CDM and FDM simulations in the quasi-linear regime. We use the speed of sound parameter of the EFT of LSS as a probe for any difference between these two models at these scales. 
The possible difference could be attributed to the back-reaction of the small-scale perturbations - sensitive to wave effects - on the quasi-linear perturbations.

It should be noted that future studies could compare the potential effects of alternative dark matter models, such as warm dark matter (WDM), which also predict the suppression of small-scale structures.

We found the speed of sound by fitting the EFT formula at one-loop order,  Eq.\eqref{eq:peft1loop}, to the power spectrum of the CDM/FDM simulations. We used CDM and FDM simulations with a size of a box of size $L= 250 h^{-1}/,\textrm{Mpc}$ and $512^3$ particles, performed using the Gadget-2 code with different initial conditions. Table. \eqref{tab:cs-compar-spt} lists the values of $c_{s(1)}^2$ derived from different simulations. The values of $c_{s(1)}^2$ for the FDM simulations completely agree with the CDM simulation within the error bars. So, we could not deduce any difference between the values of the EFT parameters for CDM and FDM models in the one-loop order.

On the other hand, though statistically insignificant, there is an interesting trend in the \textit{mean} values of the results listed in Table. \eqref{tab:cs-compar-spt}. If we compare them for different FDM masses, we see a consistently increasing trend as we decrease the mass. Since the smaller mass entails the suppression in the larger scales (smaller $k$s) and the possible deviation from the CDM should be higher, the observed trend suggests that the difference in the mean value of the speed of sound parameter for the CDM and FDM simulations is not just a statistical matter of accident, and may have a physical meaning. (To minimize the cosmic variance error due to the different realizations, we have used the same random seed numbers to generate the initial conditions for the CDM and FDM simulations.) This finding suggests that by increasing the size and resolution of the simulations and reducing the size of error bars, the minor disparity between the mean values of the CDM and FDM speed of sounds may persist, now lying beyond the error bars; that could be an interesting subject for future studies.

In addition to the relatively low resolution and size of the box, our findings also suffer from another uncertainty. We have not performed simulations solving the FDM dynamical equations. Instead, we have only used the FDM initial conditions. This is a good approximation for studying large scales. However, it is possible that if we use a simulation that incorporates the QP in its dynamics in late times, we would see a clear difference between the speed of sound values obtained for CDM and FDM simulations. Nevertheless, as discussed earlier, generating sufficiently large simulations to study quasi-linear scales is still beyond the capabilities of current FDM cosmological simulation codes. Perhaps in the future, with the development of more efficient methods or the availability of much more powerful computing resources, this goal will be within reach.

To sum up, the speed of sound parameter does not exhibit a clear difference for the CDM and FDM models in our analysis. However, the minor but interesting trend observed in the parameter across different FDM masses suggests that further investigations using more accurate and larger simulations are needed to evaluate the above conclusion.

\appendix
\section{IR limit of the Perturbations and Effective Sound Speed} \label{sec:IRlimit}
\label{1:app}

In general, for any theory, whether it is renormalizable or non-renormalizable, one must introduce a finite cut-off and add possible interactions compatible with the underlying symmetries to ensure that the predictions are independent of the cut-off. The process of renormalization is crucial in ensuring that the perturbative expansion is both physically meaningful and convergent. Within the EFT of LSS,  we extend the validity of the theory well beyond the linear regime by cutting off the theory on length scales longer than the non-linear scale. It is well-known that the perturbation theory equations hold for the scales well above the non-linear scale, $k_{NL}$. However, when going beyond linear theory, any higher-order term in perturbation expansion becomes a convolution of linear perturbations with certain kernels so that the initial fields' momenta can be "hard," i.e., close to this cut-off. In particular, this mixing leads to "loop" diagrams contributions to the correlation functions. However, since the EFT does not hold beyond some scale, the momentum integrals run up to some scale $\Lambda$. In other words, loops need to be regularized, while new interactions or "counterterms" must be included so that the final result does not depend on the cut-off. In this way, to make analytical predictions on large scales, we come up with an effective field theory characterized by several parameters, such as the speed of sound and the viscosity\cite[]{Baumann:2010tm,Carrasco:2012cv}
 
Even in the absence of long-wavelength perturbations, short-scale non-linearities, specifically the average of their dispersion velocities, $\langle v_s^2 \rangle$ give rise to an isotropic effective pressure once we focus on large-scale non-linearities. The Linear response of small-scale dispersion velocities in the presence of a long perturbation determines the effective speed of sound \cite[]{Carrasco:2012cv}. In the EFT approach to the LSS, the leading correction to the pressureless perfect fluid is the stress tensor of an imperfect fluid, characterized by the speed of sound for the fluctuations, as well as the viscosity. At the one-loop level, there is only one counterterm, namely the speed of sound $c^2_{s(1)}$. Introducing this correction not only makes analytical predictions possible but also it well quantifies different UV (short-scale) physics.

We are generally interested in the correlation function of a set of "soft" modes, whereas, in a complex Feynman diagram, some internal lines- appearing in the loops - may be hard; for clarity, we represent these hard momenta with thick lines. A time-evolution diagram is a non-stochastic contribution if all the hard lines can be paired and contracted. For instance, $\delta_3$ with the following Feynman diagram

%\nolinenumbers
\begin{equation*}
\label{Psi3}
\textrm{non-stochastic diagram:}\quad
\raisebox{-55 pt}{
\begin{tikzpicture}[scale=.7]
\begin{scope}[>=Latex,line width=2 pt]
\draw[-](0,0) -- (3,3);
\draw[-](4.5,1.5) -- (3,3);
\draw[->](0,0) -- (0.95,0.95);
\draw[->](6,0) -- (4.35,1.65);
\draw[->](0,0) -- (2.45,2.45);
\end{scope}
\begin{scope}[>=Latex,line width=1 pt]
\draw[->](3,3) -- (3,4.0);
\draw[-](3,3.74) -- (3,4.5);
\draw[->](3,0) -- (2.15,0.85);
\draw[-](3,0) -- (1.5,1.5);
\end{scope}

\draw [line width= 2 pt](0,0) -- (3.035,3.035);
\draw (0,-.5) node  {$\Q$};
\draw (3,-.5) node  {$\K$};
\draw (6,-.5) node  {$-\Q$};
\draw (3,5) node  {$\K$};

\end{tikzpicture}
}
\end{equation*}
%\linenumbers
is a non-stochastic term in the field-perturbation expansion. These contributions reflect how short-scale perturbations respond to the presence of long modes. It can be interpreted as the response - either linear or higher order - of the short-scale physics to the large-scale fluctuations.

On the other hand, when the above does not occur, it corresponds to the short-scale perturbations coincidentally aligned together to make a long mode. For example, $\delta_2$ given by the following diagram

%\nolinenumbers
\begin{equation*}
\label{Psi2}
\textrm{stochastic diagram:}\quad
\raisebox{-55 pt}{
\begin{tikzpicture}[scale=.7]
\begin{scope}[>=Latex,line width=2 pt]
\draw[->](0,0) -- (1.65,1.65);
\draw[-](1.5,1.5) -- (3,3);
\draw[->](6,0) -- (4.35,1.65);
\draw[-](4.5,1.5) -- (3,3);
\end{scope}
\begin{scope}[>=Latex,line width=1 pt]
\draw[->](3,3) -- (3,3.95);
\draw[-](3,3.74) -- (3,4.5);
\end{scope}

\draw (0,-.5) node  {$\Q$};
\draw (6,-.5) node  {$-\Q+\K$};
\draw (3,5) node  {$\K$};

\end{tikzpicture}
}
\end{equation*}
%\linenumbers

is the simplest example.
Note that one can contract \emph{all} initial hard lines in a non-stochastic diagram and then integrate over their momenta. These diagrams contribute a deterministic value to the higher-order matter perturbations. For the stochastic diagrams, however, one has to take expectation-values with many such diagrams to pair the initial lines. In this sense, these diagrams lead to in-deterministic contributions to the current value of the field fluctuations.

In particular, for the standard dark matter scenario, when the absolute value of a pair of momenta is much larger than that of the rest, we have
\begin{equation} \begin{aligned}
    F_n^{(s)}(\k_1, \dots ,\k_{n-2},\q,-\q) \propto k^2/q^2
\end{aligned} \end{equation}
where $F_{n}^{(s)}$ is the $F_n$ kernel symmetrized for its arguments, i.e. the incoming momenta, and $k^2 \equiv k_1^2+\dotsb +k_n^2$. Note that $G_n^{(s)}$ also obeys the same scaling. For a non-stochastic filed perturbation, correlated with a first-order field perturbation $\delta_1$, we get
\begin{equation} \begin{aligned}
    P_{\mathrm{non-stochastic}}(k)\propto k^2
\end{aligned} \end{equation}
Moreover, in the case of the stochastic field perturbation, where two such terms are correlated, the power scaling with the soft external momentum is $P_{\mathrm{stochastic}}\propto k^4$ \cite{Abolhasani_2016}.

\begin{acknowledgments}
We would like to express our gratitude to Mohammad Hossein Namjoo for his valuable contributions and thought-provoking discussions throughout the project. We thank Leonardo Senatore, Mehrdad Mirbabayi, and Hossein Mos'hafi for the valuable and insightful conversations. We acknowledge using the \emph{Wolfram Mathemtica} and \emph{YT-toolkit} for the data analysis and making plots. The simulations were performed using the computational resources in the High-Performance Computing Center (HPC) at the Sharif University of Technology. We also thank Alireza Baraani, who kindly helped us to use the Sharif HPC machine.
\end{acknowledgments}

\FloatBarrier
\bibliography{MAIN}{}

\end{document}